\documentclass[a4paper,11pt]{article}
\usepackage{pos}
\usepackage{slashed}

\newcommand{\Rco}{$\mathcal R_{\slashed E}$}
\newcommand{\dB}{\boldsymbol{\delta}\mathbf{B}}
\newcommand{\Bma}{\mathbf{B}}
\newcommand{\Eel}{\mathbf{E}}
\newcommand{\vE}{\boldsymbol{v_E}}
\newcommand{\uE}{\boldsymbol{u_E}}
\newcommand{\vep}{\varepsilon}
\newcommand{\Dee}{D_{\varepsilon\varepsilon}}

\title{Magnetised turbulent plasmas as high-energy particle accelerators}

\author*[a]{Martin Lemoine}

\affiliation[a]{Astroparticule et Cosmologie (APC),\\ CNRS -- Université Paris Cité,\\
  F-75013 Paris, France}

\emailAdd{mlemoine@apc.in2p3.fr}

\abstract{This proceedings paper reports on the theoretical modelling of particle acceleration in magnetised turbulent plasmas. It briefly reviews some recent findings obtained from fully kinetic numerical simulations of large-amplitude, semi to fully relativistic turbulence. The paper then argues that these findings can be understood within the framework of a ``generalised Fermi'' picture of stochastic acceleration, which it summarises. The dominant contributions to acceleration appear to arise from particle interactions with sharp, dynamic bends of the magnetic field lines and regions of velocity compression. Interestingly, the acceleration rate is spatially inhomogeneous and its probability distribution follows a broken power law extending up to large values. This makes relativistic, large-amplitude turbulence an extreme particle accelerator. Some implications for particle transport and the shape of the particle energy spectrum in the presence of radiative losses and over long timescales are also discussed.}

\ConferenceLogo{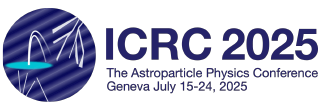}

\FullConference{39th International Cosmic Ray Conference (ICRC2025)\\
 15–24 July 2025\\
Geneva, Switzerland\\}

\begin{document}
\maketitle

\section{Introduction}
The development of large-scale, cutting-edge particle detectors and observatories in  recent decades has dramatically expanded our view of the very high-energy (VHE) multi-messenger Universe; see, e.g., the many contributions and the rapporteur talks presented at this ICRC edition. Concurrently, theorists have made substantial progress in modelling the sources of VHE photons, cosmic rays, and neutrinos, benefiting from the development of increasingly sophisticated numerical models. Yet, some critical questions persist, most notably the nature of the acceleration mechanism and its microphysics. The lack of direct observational information on the acceleration process does not help in that regard. It stems from the vast disparity between the microscopic scales of particle acceleration and the macroscopic scales governing the dynamical evolution of the source: e.g., the gyroradius of a TeV proton in a 1~G magnetic field ($r_{\rm g}\,\simeq\,3\times10^4\,$km) lies orders of magnitude below any length scale accessible to observations. Additionally, modelling particle acceleration is inherently complex because it involves nonlinear, multi-scale phenomena that often hinder direct estimates of key observables. Furthermore, particle acceleration encompasses a broad variety of mechanisms, ranging from shock acceleration and stochastic acceleration in turbulent or velocity-sheared plasmas to magnetic reconnection and electrostatic gaps in the vicinity of compact objects. 

Among these processes, stochastic acceleration stands out for its historical significance and unique characteristics. It dates back to the first concrete scenario of cosmic-ray origin, proposed by E. Fermi in two seminal papers~\cite{1949PhRv...75.1169F,1954ApJ...119....1F}. Recognising that the high conductivity of astrophysical plasmas neutralises electric fields in the plasma rest frame over relevant length and time scales, E. Fermi postulated that the accelerating electric field $\Eel$ would arise from the motion of magnetised plasma. Following ideal Ohm's law --- or alternatively a Lorentz transformation from the plasma to the laboratory frame --- $\Eel = -\vE\times \Bma/c$, where $\vE$ denotes the random velocity field, $\Bma$ the total magnetic field, including turbulent ($\dB$) and regular ($\mathbf{B_0}$) components. Particles immersed in an environment of moving inhomogeneities experience a random walk in energy space, characterised by the rate at which they encounter the randomly-oriented electric fields.

Plasma turbulence as a source of VHE particles has been proposed for a wide variety of astrophysical objects, including --- with a highly incomplete list of references --- black hole accretion discs~\cite{1996ApJ...456..106D}, galactic outflows~\cite{2011PhRvL.107i1101M}, galaxy clusters~\cite{2016MNRAS.458.2584B}, large-scale relativistic jets~\cite{2011ApJ...739...66T} and compact object mergers~\cite{2025ApJ...994L...7F}, among (many) others. Recently, it has emerged as the leading candidate for explaining the presence of $\sim 100\,$TeV protons in the black hole coronae of nearby Seyfert galaxies and the corresponding secondary neutrino signal detected by IceCube~\cite{2022Sci...378..538I}, see \cite{2020PhRvL.125a1101M} for a seminal discussion.  In phenomenological applications, stochastic acceleration is known for two key features: it provides an efficient agent of dissipation, regardless of the plasma conditions, and it tends to concentrate most of the energy into the highest energy particles, generating hard particle spectra~\cite{1984A&A...136..227S}. In contrast, relativistic shock acceleration becomes inoperative in magnetised plasmas ($\sigma\gtrsim 10^{-4}$, see below for the definition of $\sigma$)~\cite{2015SSRv..191..519S}, while reconnection generates steep spectra if the plasma is not strongly magnetised, $\sigma\lesssim 1$~\cite{2020PhPl...27h0501G}.

In this broad context, this paper reports on some recent progress in our theoretical understanding of particle acceleration in magnetised turbulent plasmas. It summarises a ``highlight talk'' given at the 39$^{\rm th}$ ICRC conference; thus, it does not aim to provide a review. Instead, it highlights some results obtained through numerical simulations (Sec.~\ref{sec:sim}), analytical developments (Sec.~\ref{sec:GF}) and some of their consequences for multi-messenger phenomenology (Sec.~\ref{sec:cons}). Within the broad parameter space of astrophysical turbulence, this paper focuses on the regime of large-amplitude ($\delta B_{\rm rms}\sim B_0$, where $\delta B_{\rm rms}\equiv\langle\delta B ^2\rangle^{1/2}$) and high eddy velocity, where most of the recent progress has been achieved in recent years. This regime is particularly favorable for particle acceleration and is thus likely representative of sources of VHE particles.

\section{Insights from kinetic numerical simulations}\label{sec:sim}
Particle acceleration in turbulent plasmas has long been studied, initially through test particle tracking in MHD simulations --- see, e.g., \cite{2023ApJ...959...28P} for a recent study, and references therein --- and more recently through fully kinetic particle-in-cell (PIC) simulations --- see, e.g., \cite{2018ApJ...867L..18Z,2019ApJ...886..122C} for pioneering studies and \cite{2025MNRAS.543.1842W} for a comprehensive recent discussion, as well as references therein. This numerical approach uniquely offers a self-consistent description of the co-evolution of charged particles and the electromagnetic fields that they induce, though this comes at the cost of resolving plasma length scales (gyroradius and skin depth), which are orders of magnitude smaller than any macroscopic length scale. Due to their high computational demand, these simulations are thus often limited to exploring the initial stages of acceleration, frequently under conditions of reduced dimensionality and artificially lowered mass ratios between species. Despite these limitations, they have provided valuable insights into particle acceleration in turbulence, including the first detailed measurements of the acceleration rate and of the non-thermal particle energy distribution.

\begin{figure}[t]
   \centering
   \includegraphics[width=0.9\columnwidth]{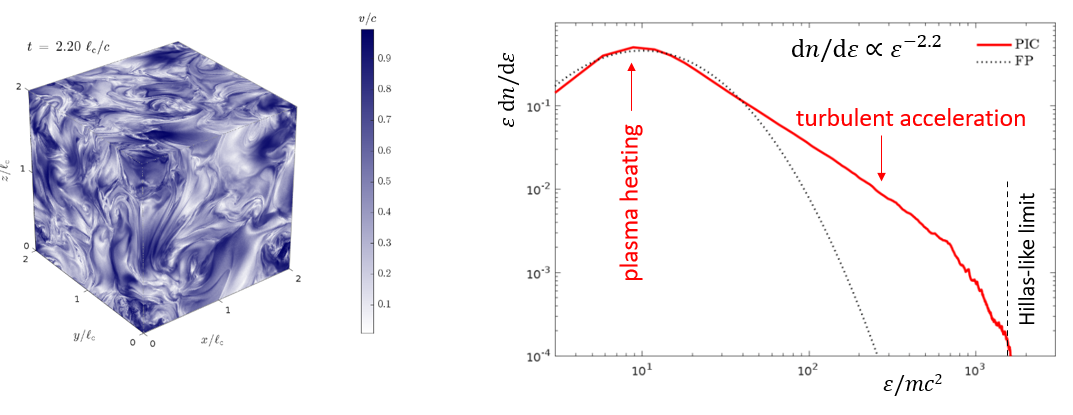}
    \caption{Illustration of a PIC numerical simulations of relativistic, collisionless, large-amplitude, driven magnetised turbulence ($\delta B_{\rm rms}/B \sim 2$, $\sigma \sim 10$). Left panel: 3D view of a simulation cube, showing the plasma velocity in units of $c$. Right panel: particle energy distribution (thick red line); the dotted black line represents the prediction from a purely diffusive Fokker-Planck model. See text for details. Adapted from \cite{2022PhRvD.106b3028B}.}
    \label{fig:PIC_spec}
\end{figure}

So far, studies have focused on semi-relativistic to fully relativistic turbulence, characterised by a rms eddy velocity $\delta v_{\rm rms}/c \sim O(1)$. Given that, on average, $ E/ B \sim \delta v_{\rm rms}/c$, the relativistic regime provides the fastest acceleration rates, and is thus more suitable to exploration via PIC simulations. Figure~\ref{fig:PIC_spec} illustrates one such simulation, for $\delta B_{\rm rms}/B_0\sim 2$ and a magnetisation level $\sigma \sim 10$, where $\sigma$ is the ratio of magnetic energy density to plasma enthalpy density.  Equivalently, $\sigma = u_{\rm A}^2/c^2$ in terms of the Alfvén four-velocity $u_{\rm A}$. A magnetisation parameter $\sigma > 1$ thus describes relativistic turbulence, since the characteristic eddy velocity $\delta v_{\rm rms} \simeq v_{\rm A}$, where $v_{\rm A}$ is the Alfvén three-velocity, in Alfvénic turbulence.

The parameter space for such simulations is rather broad, and its current exploration is far from complete. Beyond the turbulence amplitude and the eddy velocity, the turbulence may be initially imprinted as magnetic perturbations then left to decay (``decaying turbulence'') or continuously driven for a period of time, as is the case  in Fig.~\ref{fig:PIC_spec}. Particles are injected at the initial time, implying that PIC simulations provide a view of the Green's function driving the evolution of the distribution function over time. This remark may be of importance when considering astrophysical scenarios where particles are rather injected continuously in the turbulent bath. Additionally, the simulation illustrated in Fig.~\ref{fig:PIC_spec} does not include energy losses or escape losses, which can alter the spectrum~\cite{2021ApJ...921...87N,l5tb-tjb5,2025PhRvL.135f5201G}.

Nevertheless, Fig.~\ref{fig:PIC_spec} highlights the main tendencies observed so far, notably a power-law energy distribution with a thermal peak at $\vep_{\rm th}$ (here $\vep_{\rm th}\sim 10\, mc^2$),  and a cut-off at $\vep_{\rm c}$, representative of a confinement (or Hillas-like) limit where the gyroradius $r_{\rm g}(\vep_{\rm c})$ approaches the outer scale of the turbulence $\ell_{\rm c}$. This cut-off arises because once $r_{\rm g}>\ell_{\rm c}$, a particle effectively decouples from the turbulence, perceiving it as a collection of random perturbations of small extent compared to its gyroradius. In numerical simulations, $\ell_{\rm c}$ is constrained by the box size, but in real astrophysical systems, the dynamic range between the thermal peak at $\vep_{\rm th}$ and the cut-off at $\vep_{\rm c}$ can take much larger values. The spectral index $s$, defined by ${\rm d}n/{\rm d}\vep \propto \vep^{-s}$, ranges from $4$ in the sub-relativistic limit $\delta v_{\rm rms} < c$, to $3$ in the mildly relativistic limit at $\delta v_{\rm rms} \sim c$ and $2$ in the highly relativistic limit $\sigma \gg 1$, in the absence of losses~\cite{2025MNRAS.543.1842W}. If the plasma is initially cold, turbulent dissipation heats it up to $\vep_{\rm th} \sim \sigma m c^2$ when $\sigma \gtrsim 1$, as in Fig.~\ref{fig:PIC_spec}, through reconnection in microscopic current sheets~\cite{2019ApJ...886..122C}. This process also serves as a pre-injector for stochastic acceleration: once the particle becomes supra-thermal and its gyroradius increases, it decouples from the current sheets then becomes subject to stochastic acceleration in the turbulent motions on larger and larger scales.

The emergence of power-law spectra represents a rather unexpected outcome of these simulations. To date, most, if not all, studies implementing stochastic acceleration in VHE sources have utilised a purely-diffusive Fokker-Planck formalism --- meaning, solely characterised by the diffusion coefficient $\Dee$ --- that describes the evolution of the distribution function for particles undergoing simple Brownian motion in energy space. However, employing the diffusion coefficient measured in these numerical simulations, $\Dee\simeq 0.1 \sigma \vep^2 c/\ell_{\rm c}$, the corresponding solution to the Fokker-Planck equation yields a log-normal distribution that fails to reproduce the observed spectra, see the dotted line in Fig.~\ref{fig:PIC_spec}. This points to a fundamental aspect of turbulent acceleration that has been overlooked so far, and which admittedly still remains to be understood in detail. One must also note that the detailed numerical analyses conducted in \cite{2025MNRAS.543.1842W} show that a full Fokker-Planck formalism employing both advection and diffusion coefficients measured in the simulation is able to recover the measured spectra. The advection coefficient displays a non-trivial energy dependence, including negative values over some energy range, which also remains to be understood. 

One plausible interpretation for the origin of these power law spectra is that all particles are not accelerated at the same rate~\cite{2020MNRAS.499.4972L}. To see this, note that the mean acceleration time, defined by the diffusion coefficient, is $t_{\rm acc}\sim \vep^2/\Dee \sim 10 \sigma^{-1} \ell_{\rm c}/c$, thus on the order of several eddy turnover times $\ell_{\rm c}/c$ in relativistic turbulence. Since PIC simulations reveal particle energy spectra spanning several decades within just a few eddy turnover times, the highest energy particles must have been accelerated at rates significantly exceeding the mean value $\Dee/\vep^2$. In other words, the failure of the purely diffusive Fokker-Planck scheme may reflect the fact that a single mean value  ($\Dee$)  does not properly capture the complexity of the acceleration process. A strong spatial inhomogeneity of the acceleration rate has been observed in PIC simulations~\cite{2021ApJ...921...87N}, and detailed measurements presented in Sec.~\ref{sec:cons} further support the above interpretation, indicating notably that the probability distribution function (pdf) of the acceleration rate follows a broken power law, peaking near the mean and extending to large values.

Turning to theory, there is no consensual prediction for $\Dee$ and its scaling, despite extensive literature. In the original Fermi model, particles interact with randomly moving magnetic inhomogeneities in a point-like, discrete manner (the ``Fermi pinball''). At each interaction, they undergo elastic scattering by an angle of order unity in the rest frame of the scatterer. Assuming relativistic particles and $\delta v_{\rm rms} < c$, this results in
\begin{equation}
    \Dee\,\approx\, \frac{v_E^2}{c^2}\,\frac{\vep^2}{t_{\rm int}(\vep)}\,,
    \label{eq:FermiD}
\end{equation}
where $t_{\rm int}(\vep)$ denotes the characteristic interaction timescale. Since particles are deflected at each interaction, a reasonable approximation for $t_{\rm int}$ is $t_{\rm int}(\vep)\sim t_{\rm scatt}(\vep)$, the scattering time of the particle in the turbulence. However, describing the turbulence as a collection of point-like scattering centres is incomplete, as it overlooks the continuous nature of the velocity flow and the complex multi-scale nature of the fluctuations. Generalisations to more realistic continuous random flow geometries were proposed by A. Bykov and I. Toptygin at the $17^{\rm th}$ ICRC ~\cite{1983ICRC....9..313B}, further examined by V. Ptuskin~\cite{1988SvAL...14..255P}, and revisited in subsequent studies, e.g., \cite{2006ApJ...638..811C}. These works estimate the diffusion coefficient assuming that particles diffuse spatially in large scale compressive modes.

Other models attribute particle acceleration to an interplay between turbulent fluctuations and reconnection layers, building on the concept of turbulent reconnection~\cite{2020PhPl...27a2305L,2023ApJ...942...21X}. Yet, the prevailing paradigm in this field has been that of resonant wave-particle interactions, which describes the turbulence as a sum of many incoherent, small-amplitude plasma waves (e.g., Alfvén or magnetosonic), e.g.~\cite{1989ApJ...336..243S}.  Particles gyrating along the background magnetic field gain/lose energy from those waves with which they are in phase, while the cumulative effect of non-resonating waves averages out. This quasilinear description provides an analytical framework for perturbative calculations of transport coefficients. Regarding diffusion in energy space, it yields a scaling similar to Eq.~(\ref{eq:FermiD}), with the substitution $v_E \rightarrow v_{\rm A0}$ --- where $v_{\rm A0}$ is the Alfvén velocity relative to the background magnetic field --- and specifies $t_{\rm int} \simeq (\delta B_{\rm rms}/B)^{-2}\,(r_{\rm g}/\ell_{\rm c})^{2-q}\,\ell_{\rm c}/c$; the exponent $q\simeq 3/2$ or $5/3$ represents the index of the power spectrum of magnetic fluctuations. These scalings have been generalised and improved in various studies to account for the anisotropy of magnetised turbulence theories and for improved descriptions of particle transport, see e.g.~\cite{2020PhRvD.102b3003D} and references therein for an application to relativistic turbulence. Nevertheless, such quasilinear calculations rely on several assumptions. Specifically, they assume that the waves are sub-relativistic, of small amplitude, and carry random phases. This approach overlooks any form of turbulence intermittency, as any physical quantity   expressed as a sum of many incoherent waves behaves as a Gaussian random field. Consequently, these calculations cannot readily account for the broad extent of acceleration rates observed in PIC simulations.

To make progress in this direction, the following section presents an effective theory for stochastic acceleration, which directly extends the original Fermi model to a more realistic turbulent flow~\cite{2025PhRvE.112a5205L}. This theory is non-perturbative and fully covariant, and thus well-suited for addressing large-amplitude relativistic turbulence. Additionally, this framework introduces new acceleration channels that do not have a counterpart in the wave-particle interaction picture. In particular, it will be argued that the interaction of particles with sharp, dynamic bends of the magnetic field, similar to curvature drift acceleration, seemingly provides the dominant contribution to acceleration and may explain the spectra observed in PIC simulations.

\section{A generalised Fermi view}\label{sec:GF}
In the Fermi process, particles gain or lose energy in an electric field induced by the motion of magnetised plasma at velocity $\vE$ ($\Eel = -\vE\times \Bma/c$). In shock acceleration, the value of $\vE$ depends on whether the particle is upstream or downstream of the shock but is otherwise uniform in both regions. Standard practice tracks the particle trajectory in either local frame of rest and connects both frames by a Lorentz transformation.  The generalised Fermi view leverages this observation to follow the particle momentum through a sequence of instantaneous local rest frames --- denoted as \Rco\ --- defined at the instantaneous position of the particle~\cite{2019PhRvD..99h3006L}. Hereafter, the velocity of the \Rco\ frame is denoted as $\vE$, and its four-velocity as $\uE$. 

In a turbulent plasma, the flow velocity exhibits continuous spatial and temporal variations. Consequently, the particle energy evolves in \Rco\ not under the influence of $\mathbf{E}$, which vanishes there, but under that of the inertial forces associated with the variation of $\vE$ (i.e., the Fermi pinball has transformed into a rollercoaster). Within this generalised Fermi framework, the energising role of the electric field is thus supplanted by the variation of the local velocity field. This should not be surprising, as the fundamental principle of Fermi acceleration is that particles gain energy by traversing regions with differing $\vE$: if $\vE$ were uniform throughout space and time, one could perform a single boost to the globally inertial frame moving at $\vE$, in which the particle energy would remain constant. In full, the standard equation $\dot\vep = q\boldsymbol{v}\cdot\Eel$ in the laboratory frame becomes
\begin{equation}
    \frac{{\rm d}\vep'}{{\rm d}t'}\,=\,-\frac{\partial {u_E}_\beta}{\partial x^\gamma}\biggl\{
    {u_E}^\gamma{e_i}^\beta\,{p'}^i+ {e_i}^\beta {e_j}^\gamma\,{{p'}^i}{{v'}^j} \biggr\}\,.
    \label{eq:GF1}
\end{equation}
The above equation employs greek letters for spacetime indices $(0,\dots,3)$, and latin indices for the spatial components; primed quantities are expressed in the local \Rco\ frame, $p'$ refers to the particle momentum and $v'$ to its three-velocity. The quantities with mixed indices ${e_j}^\gamma$ are tetrad components linking the \Rco\ frame to the laboratory frame. Equation~(\ref{eq:GF1}) is exact and builds on the pioneering work of G. Webb~\cite{1985ApJ...296..319W} in this area.

In a turbulent plasma, the velocity $\uE$ as well as its gradients are random functions of the particle position, so that Eq.~(\ref{eq:GF1}) is in essence a stochastic differential equation. To make use of Eq.~(\ref{eq:GF1}), one must integrate a model of particle transport. This ultimately leads to an effective theory of stochastic acceleration that encompasses all forms of (non-gyroresonant)  acceleration channels, e.g., curvature drift, gradient drift, betatron, transit-time damping, magnetic pumping etc.~\cite{2025PhRvE.112a5205L}.

\begin{figure}[t]
   \centering
   \includegraphics[width=0.8\columnwidth]{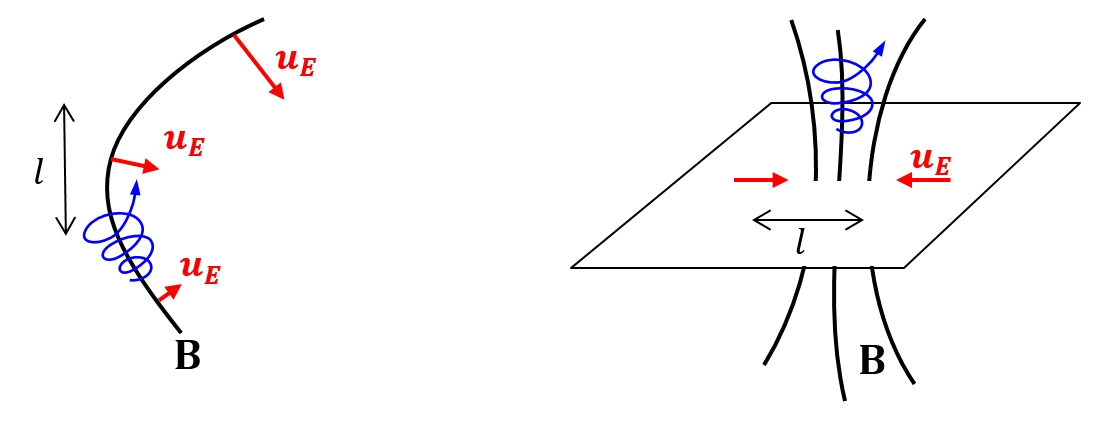}
    \caption{Sketches illustrating the mechanisms of particle energisation in a generalised Fermi scenario, for particles interacting with a perturbation of extent $l\lesssim \lambda_{\rm s}$, where $\lambda_{\rm s}$ represents the particle mean free path to scattering in the turbulence. From left to right: curvature drift, corresponding to the variation of $\uE$ along the magnetic field line; perpendicular compression of $\uE$, encompassing gradient-drift and betatron acceleration. The third channel is not depicted; it involves the acceleration of the field line projected onto the magnetic field direction. }
    \label{fig:diag}
\end{figure}

For the present purposes, it can be simplified by considering velocity perturbations on scales $l$ larger than the particle gyroradius $r_{\rm g}$ but smaller than the particle mean free path in the turbulence. By decomposing the tensor ${{u_E}^\beta}_{,\gamma}$ into its fundamental components (acceleration, shear, compression, vorticity), Eq.~(\ref{eq:GF1}) can be broken down into three contributions~\cite{2021PhRvD.104f3020L}: a term denoted ${a_E}_\parallel$ describing an effective gravity projected along the field line, associated with the acceleration of $\uE$ along the particle  trajectory; a second term denoted $\Theta_\parallel$, characterising the variation of $\uE$ along the field line, embodying curvature drift acceleration\footnote{Curvature drift is formally defined in the context of guiding centre theories for large-scale magnetic perturbations characterised by $l\gg r_{\rm g}$. The present formulation extrapolates it down to scales $l\sim r_{\rm g}$ where acceleration is most efficient. Furthermore, the mean magnetic field is here defined by coarse-graining the turbulence on a scale $r_{\rm g}$, implying that particles of different energies experience different magnetic field structures.}; and a third one written $\Theta_\perp$, describing the compression of $\uE$ in the plane transverse to magnetic field lines. Equation~(\ref{eq:GF1}) then simplifies to
\begin{equation}
    \frac{{\rm d}\vep'}{{\rm d}t'}\,=\,- p_\parallel'\,{a_E}_\parallel - p_\parallel' v_\parallel'\,\Theta_\parallel- \frac{1}{2}p_\perp'v_\perp'\Theta_\perp\,,
    \label{eq:GF2}
\end{equation}
where $\parallel$ or $\perp$ relate to the local direction of the magnetic field. The schematic diagrams illustrating these contributions to particle acceleration are shown in Fig.~\ref{fig:diag}. 

In his landmark paper, E. Fermi presented two intuitive examples of particle-scattering centre interactions:  type-A, depicting a particle bouncing off a magnetic mirror; and type-B, describing a particle navigating a bend in the magnetic field. The two contributions $\Theta_\perp$  and $\Theta_\parallel$ offer direct generalisations of these processes. The similarity is striking, yet coincidental! A crucial difference is that, in Fermi's picture, the scattering centres are static structures in their rest frames --- for example, the field line bend does not open or close --- whereas in the present case, the particle encounters a dynamically evolving structure as it traverses it. The curvature drift mechanism embodied in $\Theta_\parallel$ also provides an intuitive grasp of the underlying physics. In the laboratory frame, 
the curvature of the field line causes the particle to drift in a direction perpendicular to the plane of curvature (out of plane in Fig.~\ref{fig:diag}) along (or against) the electric field component induced by the motion of the field line in this plane of curvature. When viewed in the local comoving frame, the particle gains energy as it is propelled forward by the closing of the field line, as in a slingshot (or, more accurately, a Basque pelota chistera), but loses energy if the field line opens on itself. Another distinction from the original Fermi scheme is the origin of net energy gain. In compact form, Eq.~(\ref{eq:GF2}) can be written as $\dot\vep' \propto \vep'$, indicating that the diffusion coefficient increases with energy, thereby implying advection in energy space. The dependence of the energy gain rate on angular variables ($p'_\parallel$ and $p'_\perp$) also induces a net energy gain, by enhancing the time spent in regions where particles gain energy versus those where they lose energy. In the laboratory frame, this net energy gain results from the influence of the electric field, which focuses the spatial component of the momentum toward the direction of maximal energy gain ($q\Eel$) and defocuses it when particles move against $q\Eel$, leading to a net energy gain at order $O(E^2)$~\cite{2025PhRvE.112a5205L}. The concept of head-on and tail-on collisions loses relevance in this generalised picture.

The diagrams presented in Fig.~\ref{fig:diag} assume that the length scale $l$ of the velocity perturbation verifies $r_{\rm g}\lesssim l\lesssim \lambda_{\rm s}$. If $l>\lambda_{\rm s}$, additional acceleration channels emerge, such as turbulent shear acceleration and magnetic pumping in large-scale modes. The dominant channel likely depends on the nature of the turbulence, e.g. low or high plasma $\beta$, compressibility, etc. While this remains an open question, growing evidence suggests that the two diagrams  in Fig.~\ref{fig:diag} provide the dominant contributions in large-amplitude turbulence~\cite{2022PhRvD.106b3028B}. 

The underlying reason is contained in the pdf of field line curvature, which exhibits extended, hard powerlaw tails at large curvature in large-amplitude turbulence~\cite{2019PhPl...26g2306Y}. These tails imply that the pdf of $\Theta_\parallel$ also displays hard power-law tails, characterising strong non-Gaussianity (intermittency) of the accelerating forces. By imaging contours of $\Theta_\parallel$ or $\Theta_\perp$ in a  visualisation of a turbulence simulation~\cite{2021PhRvD.104f3020L}, one can directly demonstrate that these quantities take large values in localised regions (corresponding to the power-law tail values), implying that different particles undergo vastly different energisation histories. Some will be accelerated rapidly, while others will be accelerated slowly, depending on the regions that they encounter, as anticipated earlier in Sec.~\ref{sec:sim}. 

This observation offers a plausible explanation for the power-law energy distributions seen in PIC simulations. To illustrate this, one can develop a toy model describing the turbulence as a collection of sparse accelerating structures and solve the corresponding walk to derive the energy distribution. This approach recovers power-law spectra on intermediate time scales, which gradually converge to the lognormal form on asymptotic time scales~\cite{2021PhRvD.104f3020L}. More quantitatively, one can formulate a generalised transport equation that takes into account the full pdf of the random forces acting on the particle~\cite{2022PhRvL.129u5101L}. By measuring the pdf of these random forces in a numerical MHD simulation and combining them with this transport equation to derive the time-dependent particle distribution in energy space, one obtains power-law energy distributions. Furthermore, these distributions align well with the spectra obtained by Monte Carlo particle tracking using the full Lorentz force in the same MHD simulation~\cite{2022PhRvL.129u5101L}. The commonly-used Fokker-Planck equation is an approximation of this generalised transport equation, applicable when jumps in energy space are small and occur over short time scales. However, in large-amplitude turbulence, these jumps can become substantial, as can the time required to complete them.

This generalised Fermi model offers a new framework for understanding particle acceleration in turbulent plasmas, at the price of requiring specific knowledge of the pdf of the energising forces, which are expected to depend on the ambient physical conditions. A key open question is whether this model also accounts for particle acceleration at low amplitudes ($\delta B_{\rm rms}\ll B$), or if the quasilinear picture of randomly phased waves is restored in this limit. 

\section{Some phenomenological consequences}\label{sec:cons}
\subsection{Particle transport}
Thus far, the discussion has centered on particle acceleration; however, turbulence also mediates spatial diffusion through scattering on magnetic field inhomogeneities, even in the magnetostatic limit $\delta v_{\rm rms}/c\rightarrow 0$. A direct corollary of the preceding discussion is that particle interactions with sharp curvature bends of the magnetic field lines, or regions of strong compression, can ensure scattering. In magnetospheric plasma physics, it is known that a particle of gyroradius $r_{\rm g}$ interacting with a curvature bend on a scale $l\gtrsim r_{\rm g}$ suffers an instantaneous large-angle scattering event if the curvature $\kappa_l$ of the field line (i.e., the inverse curvature radius) satisfies $\kappa_l r_{\rm g}\gtrsim 1$. While large-scale mirroring events reverse the pitch-angle while preserving the magnetic moment, this resonant curvature scattering is nonadiabatic, leading to a violation of the magnetic moment of order unity, and thus genuine scattering and pitch-angle randomisation. 

Interestingly, it can be shown that in large-amplitude MHD turbulence, the filling fraction of regions with the necessary properties, namely $l\gtrsim r_{\rm g}$ and $\kappa_l r_{\rm g}\gtrsim 1$, is sufficient to sustain spatial diffusion~\cite{2023JPlPh..89e1701L,2023MNRAS.525.4985K}. This result arises from the power-law distribution of $\kappa_l$ at large values, which ensures that, regardless of $r_{\rm g}$, regions with sufficiently large $\kappa_l$ to induce resonant curvature scattering seem to exist in sufficient numbers. Measurements of the statistics of $\kappa_l$ as a function of the scale $l$ indicate that the corresponding filling fraction $f_l$ evolves as a power law in terms of $l$, and hence $r_{\rm g}$ when considering $l\sim r_{\rm g}$, and therefore particle energy. A simple estimate of the particle mean free path, $\lambda_\parallel \simeq l/f_l$ at $l\sim r_{\rm g}$, then yields $\lambda_\parallel \approx (r_{\rm g}/\ell_{\rm c})^\alpha\,\ell_{\rm c}$ with $\alpha \simeq 0.3 - 0.5$. This result, highly relevant to cosmic-ray phenomenology, suggests that localised sharp bends of the magnetic field lines may be the dominant agent of particle scattering in large-amplitude turbulence, offering new ways to understand particle transport; see also~\cite{2025arXiv250518155L,2025ApJ...988..269B}. Further studies remain mandatory to better understand the underlying statistics and their dependence as a function of turbulence amplitude and the ratio $r_{\rm g}/\ell_{\rm c}$.

\subsection{Impact of radiative losses on acceleration}
In practical applications, one determines the maximum energy $\vep_{\rm max}$ by comparing the acceleration rate with the radiative loss rate $\nu_{\rm syn}$. The mean acceleration rate is given by $\overline \nu_{\rm acc} = 4\Dee/\vep^2 \sim 0.4 \sigma c/\ell_{\rm c}$ (assuming $\delta B_{\rm rms}\sim B$). As mentioned in Sec.~\ref{sec:sim}, this implies that it takes several large-scale eddy turnover times on average to accelerate a particle, so that turbulent acceleration is not, in a first approximation, a rapid process. Alternatively, one can compare $\overline\nu_{\rm acc}$ to the optimal Bohm acceleration rate $\nu_{\rm acc}^{\rm B}\equiv c/r_{\rm g}$, for which the particle energy roughly doubles in a gyrotime. This comparison gives $\overline \nu_{\rm acc}/\nu_{\rm acc}^{\rm B} \sim 0.4\sigma (r_{\rm g}/\ell_{\rm c})\ll1 $ at  $r_{\rm g}\ll\ell_{\rm c}$. 

However, this assumes that $\overline\nu_{\rm acc}$ adequately represents the acceleration rate for the entire particle population, which is not the case in large-amplitude turbulence. Instead, the pdf of the acceleration rate rather follows a broken power-law shape, peaking near $\overline\nu_{\rm acc}$, and extending up to $\nu_{\rm acc}^{\rm B}$ with ${\rm pdf}(\nu_{\rm acc})\propto \nu_{\rm acc}^{-2}$. This scaling has been measured in a PIC simulation that includes synchrotron radiative losses, with $\delta B_{\rm rms}/B\sim 1$ and $\sigma\sim1$~\cite{l5tb-tjb5}. Based on this, one does not expect to observe a cut-off at the maximum energy  $\vep_D$ predicted by comparing $\overline \nu_{\rm acc}$ with $\nu_{\rm loss}$ but rather a power-law extension reflecting the distribution of acceleration rates beyond the mean value. This  spectral shape is indeed observed in the same PIC simulation. Within the constraints of these simulations, namely the finite dynamic range and the need to rescale the magnitude of losses to observe their effect within the simulation duration, the particle energy distribution is seen to turn over around $\vep_D$, displaying a spectral slope $s\simeq 4$ instead of $s\simeq 3$ in the absence of losses. Interestingly, this spectrum extends up to the maximal possible values, either $\vep_{\rm c}$ (confinement limit at $r_{\rm g}\simeq \ell_{\rm c}$) or $\vep_{\rm rad}$, the synchrotron burn-off limit where particles cool within a gyro-time. The spectrum also exhibits substantial time variability, especially as the energy approaches $\vep_{\rm rad}$. This variability reflects the intermittent nature of regions with high acceleration rate: the higher the local acceleration rate, the rarer these regions become. Naturally, when integrating over a region of extent $L>\ell_{\rm c}$, the overall variability will decrease in proportion to $(L/\ell_{\rm c})^{-3/2}$. 

These findings have direct phenomenological implications. Notably, they suggest that relativistic turbulence acts as an extreme accelerator, capable of accelerating particles at the Bohm rate and displaying a distinct signature through its time variability. This is particularly relevant for variable sources such as blazars and pulsar wind nebulae, especially the Crab Nebula, which exhibits both extreme acceleration and flaring behavior at the highest energies~\cite{l5tb-tjb5}.

\subsection{Long-term evolution}
As mentioned earlier, PIC simulations are inherently limited in duration and scale, so that one cannot always directly apply their results to astrophysical objects. The long-term evolution, in particular, deserves careful scrutiny, as particle acceleration consumes turbulent energy, potentially to the point where the turbulent cascade is altered, thereby modifying in turn the acceleration rate~\cite{1979ApJ...229..413E,1979ApJ...230..373E}. To estimate the point at which feedback becomes significant, one can compare the rate $u_p \overline\nu_{\rm acc}$ at which particles draw energy from the cascade with the rate $u_B\, \delta v_{\rm rms}/\ell_{\rm c}$ at which energy flows through the cascade, from large to small spatial scales; here $u_p$ and $u_B$ respectively denote the particle and turbulent energy densities. Once these rates become comparable, the turbulence cascade damps below some characteristic length scale $l_{\rm damp}$, slowing down the acceleration of particles whose gyroradius $r_{\rm g}\ll l_{\rm damp}$~\cite{2024PhRvD.109f3006L}. Since $\overline\nu_{\rm acc}\simeq 0.4 \sigma c/\ell_{\rm c}$, damping becomes significant once $u_p/u_B \sim 0.4 \sigma \delta v_{\rm rms}/c$, i.e., once near equipartition $u_p/u_B \sim O(1)$ is reached in mildly relativistic turbulence.

When backreaction becomes significant, stochastic acceleration enters a nonlinear phase and becomes self-regulated by the impact of high-energy particles on the acceleration process via turbulence damping. A generic consequence is the remodelling of the particle distribution, resulting in ${\rm d} n/{\rm d}\vep \propto \vep^{-2}$ at high energies ($r_{\rm g}\gtrsim l_{\rm damp}$). The global particle distribution then adopts a broken power-law form, with equal energy per decade above the break, consistent with the equipartition argument. While these results appear generic in the long-term limit, they may be modulated by turbulent dissipation through plasma heating, by the influence of energy losses or escape from the region~\cite{2025PhRvL.135f5201G}. Up to such effects, it predicts a universal spectral shape capping the spectrum by energetic arguments, regardless of the detailed microphysics of particle acceleration. In fact, this spectral shape may explain the ${\rm d}n/{\rm d}\vep\propto \vep^{-2}$ spectra observed in PIC simulations at large magnetisation ($\sigma \gg 1$), where turbulent dissipation into high-energy particles is highly efficient.

Finally, it may also have direct phenomenological consequences in a number of astrophysical sources. One notable example is the acceleration of protons to $\sim 100\,$TeV in the turbulent coronae of supermassive black holes, motivated by the recent IceCube detection of $\sim 10\,$TeV neutrinos from nearby Seyfert galaxies~\cite{2022Sci...378..538I,2020PhRvL.125a1101M}. In the case of NGC~1068, the proton luminosity inferred from the neutrino flux is so substantial ($> 1\,$\% of Eddington) that some form of equipartition with the turbulent pressure has likely been achieved. A detailed examination indicates that self-regulated stochastic acceleration can indeed produce the required spectral shape and altogether account for the high proton luminosity~\cite{2025A&A...697A.124L}.

\section{Conclusions}
Our understanding of stochastic particle acceleration has advanced significantly in recent years, particularly due to insights gained from large-scale numerical simulations. A key finding is that these simulations consistently exhibit power-law energy distributions for particles, contrary
to the expected lognormal forms based on the scaling of the diffusion coefficient $\Dee\propto \vep^2$. A plausible interpretation, supported by studies reviewed here, suggests that the acceleration process is not adequately characterised by its mean acceleration rate (or diffusion coefficient). Instead, the acceleration rate varies widely across different regions of the turbulent plasma, with its pdf following a broken power law form. This pdf peaks near the mean value $\overline\nu_{\rm acc}$ that characterises $\Dee$, and extends well beyond. This explains, at least qualitatively, why PIC simulations can observe extended particle  distributions spanning several orders of magnitude within a few large-scale eddy turnover times, even though the characteristic acceleration time inferred from $\Dee$ is itself a few eddy turnover times.

This paper has also summarised efforts in building a ``generalised Fermi'' picture of stochastic acceleration, extending  the seminal ideas of E. Fermi to a realistic turbulent setting. In this generalised Fermi framework, the dominant acceleration channel arises from the interaction of  particles with dynamic curved magnetic field lines, similar to curvature drift acceleration. The recent observation that the pdf of curvature of magnetic field lines exhibits a non-Gaussian distribution with hard power-law tails in large-amplitude turbulence offers a reasonable explanation for the distribution of acceleration rates mentioned earlier. Incorporating the full extent of this pdf in a generalised transport equation confirms the ability to reproduce the particle spectra obtained through Monte Carlo particle tracking. This provides a novel perspective on stochastic particle acceleration, wherein particles gain energy when crossing a field line bend that is closing onto itself and lose energy when crossing another that is opening on itself. 

Finally, several consequences and applications have been reviewed. Firstly, interactions with sharp magnetic bends can sustain spatial diffusion, even in less dynamic or magnetostatic turbulence. Secondly, the extended distribution of acceleration rates implies that in the presence of energy losses the spectrum does not abruptly cut off at the energy typically inferred by comparing the mean acceleration rate with the radiative loss rate. Instead, it extends beyond this point with a steepened slope, all the way up to the synchrotron burn-off limit (in relativistic turbulence). Lastly, in the long-term limit, as the turbulence cascade is dissipated into VHE particles, the acceleration process becomes self-regulated. This remodels the particle spectrum into a near universal broken power-law shape, with ${\rm d}n/{\rm d}\vep\propto \vep^{-2}$ at high energies, subject to the influence of energy losses and escape.

Without doubt, many questions deserve further scrutiny, including the microphysics of the acceleration process in lower-amplitude turbulence ($\delta B_{\rm rms} < B_0$), where the properties of intermittency may differ, and the physics of acceleration deep within the inertial range where $r_{\rm g}\ll\ell_{\rm c}$, which has so far been challenging to access in PIC simulations.

\acknowledgments{It is a pleasure to acknowledge discussions and collaborations with S. Aerdker, A.~Bykov, L.~Comisso, S. Le Bihan, M.~Malkov, K.~Murase, F.~Rieger, L.~Sironi, D. Uzdensky, V. Zhdankin and more particularly V.~Bresci, C.~Demidem, L.~Gremillet, G.~Pelletier, and A.~Vanthieghem. The research reported here has benefited from computer and storage resources by GENCI at TGCC thanks to the grants 2023-A0160411422, 2024-A0160411422 and 2025-A0160411422 on the Joliot-Curie supercomputer on the ROME partition. }

\bibliographystyle{JHEP}
\bibliography{refs}


\end{document}